\documentclass[%
reprint,
twocolumn,
superscriptaddress,
showpacs,
preprintnumbers,
graphicx,
floatfix,
byrevtex,
amsfonts,
amsmath,
amssymb,
aps,
prb,
]{revtex4-1}

\usepackage[]{graphicx}
\usepackage{epstopdf}
\usepackage{MnSymbol}
\usepackage{latexsym}
\usepackage{color}

\usepackage{dcolumn}
\usepackage{bm}
\usepackage[mathlines]{lineno}

\begin{document}

\title{Coherent Dirac-plasmons in topological insulators}

\author{Richarj Mondal}
\email{rmondal@bk.tsukuba.ac.jp}
\affiliation{Division of Applied Physics, Faculty of Pure and Applied Sciences, University of Tsukuba, 1-1-1 Tennodai, Tsukuba 305-8573, Japan}

\author{Akira Arai}
\affiliation{Division of Applied Physics, Faculty of Pure and Applied Sciences, University of Tsukuba, 1-1-1 Tennodai, Tsukuba 305-8573, Japan}

\author{Yuta Saito}
\affiliation{Nanoelectronics Research Institute, National Institute of Advanced Industrial Science and Technology (AIST), Tsukuba Central 5, 1-1-1 Higashi, Tsukuba 305-8565, Japan}

\author{Paul Fons}
\affiliation{Nanoelectronics Research Institute, National Institute of Advanced Industrial Science and Technology (AIST), Tsukuba Central 5, 1-1-1 Higashi, Tsukuba 305-8565, Japan}

\author{Alexander V. Kolobov}
\affiliation{Nanoelectronics Research Institute, National Institute of Advanced Industrial Science and Technology (AIST), Tsukuba Central 5, 1-1-1 Higashi, Tsukuba 305-8565, Japan}

\author{Junji Tominaga}
\affiliation{Nanoelectronics Research Institute, National Institute of Advanced Industrial Science and Technology (AIST), Tsukuba Central 5, 1-1-1 Higashi, Tsukuba 305-8565, Japan}

\author{Muneaki Hase}
\email{mhase@bk.tsukuba.ac.jp }
\affiliation{Division of Applied Physics, Faculty of Pure and Applied Sciences, University of Tsukuba, 1-1-1 Tennodai, Tsukuba 305-8573, Japan}

\begin{abstract}
We explore the ultrafast reflectivity response from photo-generated coupled phonon-surface Dirac plasmons in Sb$_2$Te$_3$ topological insulators several quintuple layers thick. The transient coherent phonon spectra obtained at different time frames exhibit a Fano-like asymmetric line shape of the $A^2_{1g}$ mode, which is attributed to quantum interference between continuum-like coherent Dirac-plasmons and phonons. By analyzing the time-dependent asymmetric line shape using the two-temperature model (TTM), it was determined that a Fano-like resonance persisted up to $\approx$1 ps after photo-excitation with a relaxation profile dominated by Gaussian decay at $\leq$200 fs. The asymmetry parameter could be well described by the TTM for $\geq$200 fs, therefore suggesting the coherence time of the Dirac plasmon is $\approx$200 fs. 

\end{abstract}

\date{\today}

\pacs{78.47.jg, 63.22.-m, 63.20.kd}

\maketitle

Coherent states in condensed media are quantum mechanically described in terms of the annihilation operator, as described by Glauber \cite{Glauber,Zhang90}. The concept of coherent states has succeeded in producing the {\it laser} \cite{Schawlow58}. In the last two decades, moreover, the importance of coherent states in solid state physics has greatly increased and new physical effects, such as Bose-Einstein condensation have been discovered \cite{Anderson95}.
In addition, spin relaxation in quantum spin Hall (QSH) systems \cite{Fu07} has been often studied coupled with the recent discovery of topological insulators \cite{Fu07,Zhang09}, and the loss of coherence  exhibited by its Gaussian relaxation behavior \cite{Dobrovitski08,Yang17}, a characteristic significantly different from exponential relaxation. Thus, Gaussian relaxation is often referred to as a signature of a coherent state. 

A topological insulator (TI) \cite{Zhang09} is a quantum electronic material, which is characterized by an insulating gap in the bulk, while gapless surface states (SSs) exist at the interface with the vacuum or other dielectric materials. The metallic surface states are characterized by massless Dirac quasiparticles, whose scattering is prohibited by time reversal symmetry \cite{Moore}. Exploiting the birth and decay of quasiparticles on the surface of TIs provides a novel paradigm for future application of TIs to quantum computation \cite{Kitaev}, spin electronics \cite{Chen}, and optical devices \cite{Limonov}. The dynamics of the quasiparticles on a TI surface, however, have been exclusively investigated by means of time- and angle-resolved photoemission spectroscopy under vacuum conditions and only limited information has been obtained \cite{Sobota12}. 

Recently Dirac plasmons have been observed on the surface of TIs in the form of a polariton wave using metamaterials (MMs) \cite{Pietro}. Without the help of MMs, in general, optical techniques are unable to characterize the dynamics of Dirac plasmon-polaritons on the surface of TIs, since momentum conservation requires large wavevectors for the plasmon wave to couple to a photon. Instead of using MMs, one can excite Dirac plasmons by direct coupling using a $\approx$1.5 eV photon to access the TI surface states \cite{Glinka}. In this alternative case, a film thinner than $\sim$15 nm for a Bi$_2$Se$_3$ TI is required. Under the conditions that both the Dirac plasmon, which acts as a continuum state, and surface phonons, which act as discrete states, are coherently excited, a Fano-like resonance can occur, resulting in asymmetric phonon spectra \cite{Fano,Zhang11}. 
Because of the possible plasmonic enhancement of electron-phonon coupling in the Dirac surface states for ultrathin ($\leq$10 nm) TIs \cite{Glinka15} and a possible enhancement due to Fr\"{o}hlich interactions for polar-optic modes \cite{Zhang11,Heid}, one may observe dynamical Fano-like resonances \cite{Yoshino,Watanabe17} in the vicinity of the interface between the bulk and surface regions [inset of Fig. 1(a)]. 

In this article, we have explored the ultrafast dynamics of coupling between Dirac plasmons and coherent optical phonons in ultrathin TI films, i.e. several quintuple layers thick $p$-type Sb$_2$Te$_3$, using an optical pump-probe method combined with a time-frequency analysis. The coherent phonon spectra of the intra-layer optical phonon mode exhibits a time-dependent Fano-like asymmetric line shape. The inverse of the asymmetry parameter obtained at different times for Sb$_2$Te$_3$ was not constant, but was found to strongly decay with a decay profile that could be fit to a Gaussian function for the initial time scale of $\leq$200 fs, 
after which it decayed within 1 ps with a time dependence that could be well fit to the electron temperature calculated by the two-temperature model. 
We argue that the observed Gaussian decay is a consequence of the coherence time of the Dirac plasmon (continuum state), which couples with a discrete phonon state. 
\begin{figure}
\includegraphics[width=8.5cm]{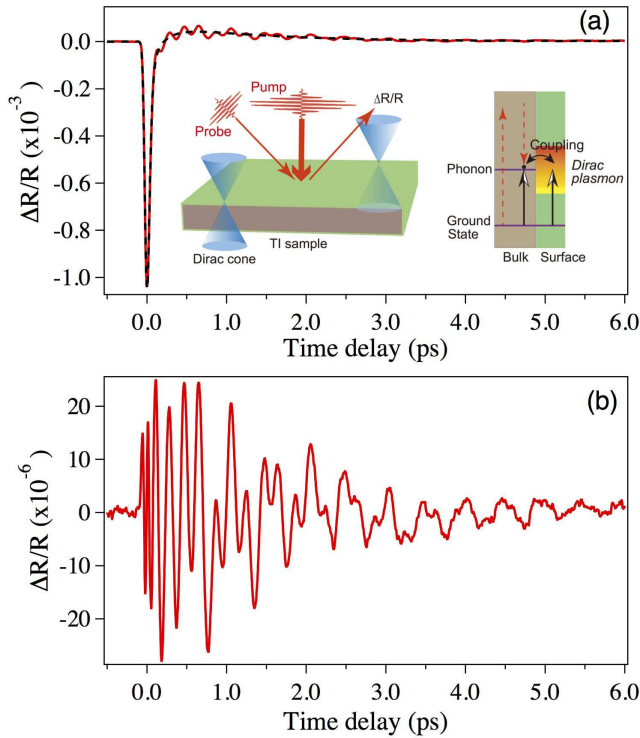}
\caption{(Color online) (a) Time-resolved reflectivity signal observed for a Sb$_2$Te$_3$ 8 QL-thick film. The dashed line represents the background. The inset represents a schematic of the experimental layout and the coupling dynamics in the surface region. (b) The time evolution of the coherent phonon oscillation signals obtained after background subtraction.
}
\label{Fig1}
\end{figure}
In the present study, we have used rectangular pieces of highly oriented polycrystalline Sb$_2$Te$_3$ 8 quintuple-layers (QL) (8 nm) thick films \cite{note1}, which were grown on Si (100) substrates by self-organized van der Waals epitaxy using helicon-wave radio frequency magnetron sputtering \cite{Saito,note2}. To prevent oxidation, samples were capped by a 20 nm-thick ZnS-SiO$_{2}$ layer. 
The ZnS-SiO$_{2}$ capping layer has long been used as a dielectric layer in optical disc applications \cite{Ohshima96} and it is completely transparent to near infrared light and makes no contribution to the observed signal.
Optical pump-probe measurements were carried out using a femtosecond Ti:sapphire laser oscillator operated at 80 MHz, which provides near infrared optical pulses with a pulse duration of $\leq$ 30 fs with a central wavelength of 830 nm. The average power of the pump beam was maintained at 120 mW. 
The $s$-polarized pump and the $p$-polarized probe beam were co-focused onto the sample to a spot size of about 70 $\mu$m with an incident angle of about $15^{\circ}$ and $10^{\circ}$ with respect to the sample normal, respectively. 
The optical penetration depth at 830 nm was estimated from the absorption coefficient to be $\approx$14 nm, which is larger than the sample thickness ($\approx$8 nm). Thus the optical excitation was homogeneous over the entire sample thickness, and the effects of the penetration depth do not play a role on the observed Fano-like asymmetric line shape in the present study. The delay between the pump and the probe pulses was scanned by an oscillating retroreflector operated at a frequency of 19.5 Hz up to 15 ps \cite{Hase12,Hase15}. The measurements were performed in air at room temperature. 

The measured time-resolved reflectivity signal ($\Delta R/R$) as shown in Fig. 1(a) consists of oscillatory components with a non-oscillatory background. The oscillatory components are attributed to coherent phonon modes while the non-oscillatory component is related to the excitation and relaxation of nonequilibrium electrons and the subsequent lattice heating \cite{Misochko16}. To obtain the oscillatory signal, the non-oscillatory components were subtracted from the transient reflectivity signal by fitting the data to a linear combination of exponentially decaying functions \cite{Misochko16}, as shown by the dashed line in Fig. 1(a). The residual oscillatory signal shown in Fig. 1(b) is assigned to coherent phonons, whose wavevector satisfies $k$ $\approx$ 0 \cite{Zeiger}. The time evolution of the oscillatory pattern is modulated by various frequency components resulting in complex behavior in the time domain. In order to uncover the modulated oscillatory signal, we utilized the discrete wavelet transform (DWT) \cite{Hase02,Hase03} and  fast Fourier transforms (FFT). The analysis, based on the DWT, was carried out using a Gabor wavelet function given by a Gaussian function \cite{Combes89}. 
Here we have used a time window of 0.54 ps corresponding to three oscillation periods of the $A^2_{1g}$ mode. Figure 2(a) presents the time-frequency chronograms obtained by DWT, together with the static FFT spectra in Fig. 2(b), obtained from the residual oscillatory signal [Fig. 1(b)]. Both the DWT chronograms and the FFT spectra exhibit two dominant phonon peaks at the frequencies of 2.05 and 5.04 THz in addition to a relatively low intensity peak at 3.33 THz. These three phonon modes are assigned to the $A^1_{1g}$, $A^2_{1g}$, and $E^2_{g}$ phonon modes, respectively, as indicated in Fig. 2(b). The peak frequencies are consistent with the values reported for transient reflectivity \cite{Norimatsu} and Raman scattering measurements \cite{Chis}. 

\begin{figure}
\includegraphics[width=8.5cm]{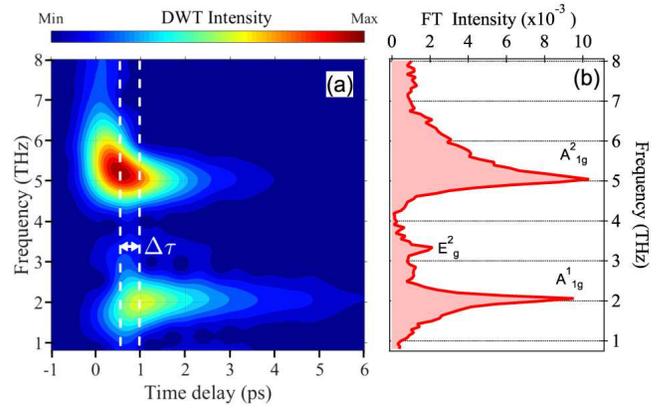}
\caption{(Color online) (a) DWT chronogram obtained from the oscillatory signal shown in Fig. 1(b). The color bar represents the DWT intensity. A time lag between the $A^1_{1g}$ and $A^2_{1g}$ modes can be clearly observed as indicated by $\Delta\tau$. (b) Static FFT spectrum obtained from the oscillatory signal in Fig. 1(b).
}
\label{Fig2}
\end{figure}

\begin{figure}
\includegraphics[width=8.5cm]{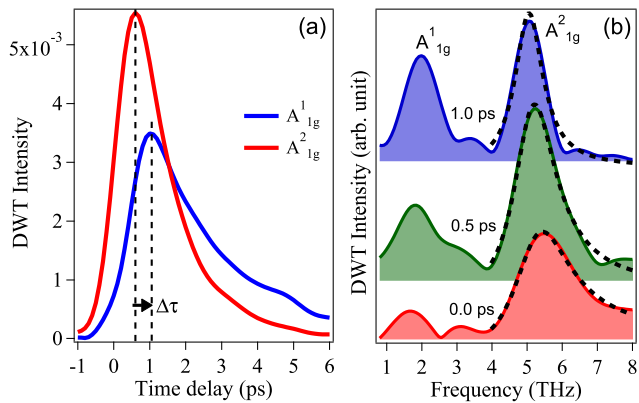}
\caption{(Color online) (a) Time evolution of the DWT intensity for the $A^1_{1g}$ and $A^2_{1g}$ modes as a function of the time delay. The vertical dashed lines represent the peak positions. (b) Time sliced DWT phonon spectra obtained for the different time delays at 0.0, 0.5, and 1.0 ps. The dashed curves indicate the fit to the spectra using Eq. (1). For clarity, the spectra have been vertically shifted. 
}
\label{Fig3}
\end{figure}

To gain new insight into these DWT chronograms, we have further investigated the $A^1_{1g}$ and $A^2_{1g}$ modes in the time-frequency domain as described below. Figure 3(a) presents the time evolution of the DWT intensity for the $A^1_{1g}$ and $A^2_{1g}$ modes, obtained from the DWT chronogram in Fig. 2(a). The intensity of the $A^2_{1g}$ mode is stronger than that of the $A^1_{1g}$ mode and decays with a time constant of $\tau_{A2}$ $\approx$1.13 ps (obtained from an exponential fit), whereas the $A^1_{1g}$ mode decays more slowly with a time constant of $\tau_{A1}$ $\approx$2.79 ps. The lifetime of the $A^2_{1g}$ mode is much shorter than the $A^1_{1g}$ mode, a result also found in previous studies on Bi$_2$Te$_3$ \cite{Wang,Misochko16,Misochko15}, Sb$_2$Te$_3$ \cite{Norimatsu}, and Bi$_2$Se$_3$ \cite{Nakamura,Norimatsu13}. 
In our experiment, however, a new aspect of coherent phonon generation is uncovered; the two modes reach a maximum intensity with different time constants and the time lag ($\Delta\tau$) between the two maximums of the $A^2_{1g}$ and $A^1_{1g}$ modes was found to be $\approx$0.5 ps, as indicated in Figs. 2(a) and 3(a). 

Surprisingly, no significant time lag ($\Delta\tau$) between the two modes was found for the 3 QL thick-film as well as for bulk (50 QL-thick) samples 
(see Supplemental Material \cite{Sup}). Here, differences in the coherent phonon generation mechanism could be a possible reason for the time lag between these two modes in the thin film TI (8 QL). In general,  coherent phonons are generated by optical excitation via either impulsive stimulated Raman scattering (ISRS) \cite{Yan85} or the displacive excitation of coherent phonon (DECP) \cite{Zeiger} processes. 
Besides these mechanisms, a coherent phonon can also be generated by a thermal gradient force (an electron temperature gradient), which can be considered to be a refined version of DECP \cite{Wang}. 
In the case of the 50 QL-thick TI, however, the thermal gradient force can only give rise to simultaneous individual atomic motion, as a consequence, there would be no time lag between the generation of the $A^2_{1g}$ and $A^1_{1g}$ modes \cite{Wang}. Whereas, for an 8 QL TI, whose thickness is less than the optical penetration depth ($\approx$14 nm), optical excitation results in a nearly uniform thermal gradient. Thus, a uniform thermal force would not produce coherent phonons in the 8 QL sample. 
We argue that a difference in the initial phase, e.g., $\sin\Omega t$ or $\cos\Omega t$, cannot explain the time lag of $\approx$0.5 ps, since the initial phase difference can only lead to a 1/4 of a phononic cycle delay, that is 0.12 ps. 
Alternatively, we propose that a different mechanism consisting of a thermal gradient force due to the electron temperature $T_{e}$ and the lattice temperature $T_{L}$ (described later), can affect the generation of the $A^1_{1g}$ mode. 

It is interesting to note that an explicit asymmetric line shape on the higher frequency side of the $A^2_{1g}$ modes is clearly observed up to $\approx$1.0 ps in the DWT chronograms [Fig. 2(a)] and in the static FFT spectra [Fig. 2(b)]. Such an explicit asymmetric line shape for the $A^2_{1g}$ mode has not been reported for bulk Sb$_2$Te$_3$ using time-resolved transient reflectivity measurements \cite{Norimatsu}, but has been seen in Bi$_2$Te$_3$ although a dynamical Fano-like resonance was not observed \cite{Misochko15}. Nevertheless, the asymmetric line shape of the $A^2_{1g}$ mode observed in our thin Sb$_2$Te$_3$ film can be attributed to quantum interference between discrete and continuum states or Fano-resonances \cite{Fano,Zhang11}. 
Note that a Fano-like asymmetric line shape is not observed for a very thin film (3 QL) (see Supplemental Material \cite{Sup}), as it is a trivial 2D insulator \cite{Jiang12,Mondal18}, in which the linearly dispersed Dirac SSs are absent. 
In addition, the $A^2_{1g}$ phonon mode in a bulk (50 QL thick) sample exhibits a symmetric line shape (see Supplemental Material \cite{Sup}). 
Taking the optical penetration depth ($\approx$14 nm) into account, the contribution from plasmons in the bulk region (normal insulator) dominates the electron-phonon coupling in the 50 QL-thick sample. Therefore, absence of the Fano-like asymmetric line shape in the bulk sample suggests that bulk plasmons do not play a significant role in the present study. 
Consequently, the observed results indicate that the Fano-like line shape in the TI has a surface origin. Similar observations have been made in (Bi$_{1-x}$In$_x$)Se$_3$ TI, referring to the spatial overlap between the surface continuum-like Dirac plasmon wave function and the discrete bulk phonon state \cite{Sim}.  

Figure 3(b) displays sliced DWT phonon spectra for three different time frames, corresponding to the time delays  0.0, 0.5, and 1.0 ps, respectively. The $A^2_{1g}$ phonon mode exhibits a strongly asymmetric Fano-like line shape at 0.0 ps, while with increasing delay, the Fano-like line shape changes to a Lorentzian line shape at 1.0 ps. To understand the time evolution of the asymmetric nature of the phonon spectra, we fit the Fano function to the experimental DWT phonon spectra \cite{Fano,Watanabe17}:
\begin{equation} \label{eq1}
I(\varepsilon,q)=\frac{(q+\varepsilon)^2}{1+\varepsilon^2},
\end{equation}
where $\varepsilon=(\omega-\omega_0-\Delta\omega_p)/\Gamma_p$, $\omega_0$ is the unperturbed phonon frequency, $\Delta\omega_p$ is the frequency shift (the real part of the phonon self-energy), $\Gamma_p$ is the line width parameter related to the phonon lifetime (the imaginary part of the phonon self-energy), and $q$ is the asymmetry parameter. 
For convenience, $q$ is represented by 1/$q$, which serves as a measure of the coupling strength. A large 1/$q$ value can be attributed to a strong interaction, and leads to the Fano line shape, whereas a small 1/$q$ value indicates a negligibly small interaction, which leads to a Lorentzian line shape. The delay dependence of the $A^2_{1g}$ phonon spectra could be well fit with the Fano function as shown in Fig. 3(b).

\begin{figure}
\includegraphics[width=8.0cm]{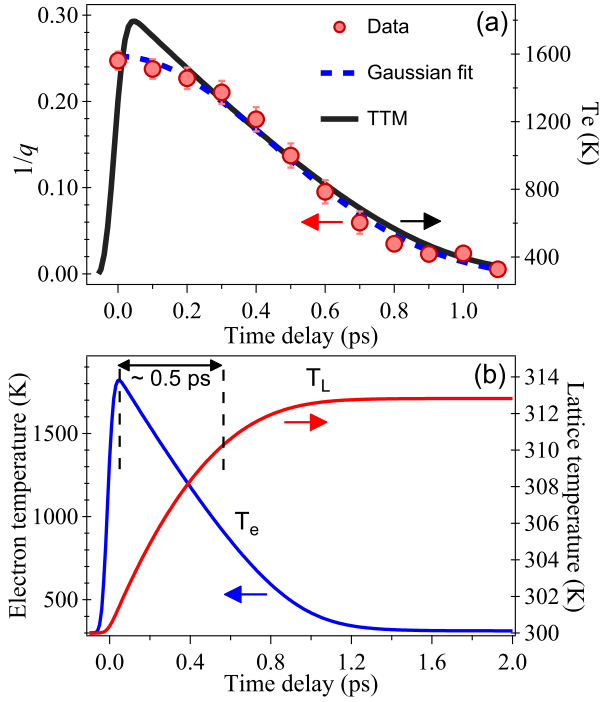}
\caption{(Color online) 
(a) The time evolution of the coupling strength (1/$q$). The solid line represents a fit using the TTM using the $e-ph$ coupling constant of $G$ = 2.0$\times$10$^{16}$ Wm$^{-3}$K$^{-1}$, whereas the dashed line represents a Gaussian fit. (b) The time evolution of the electron and lattice temperatures for Sb$_2$Te$_3$ calculated by the TTM. 
}
\label{Fig4}
\end{figure}

\begin{figure}
\includegraphics[width=8.5cm]{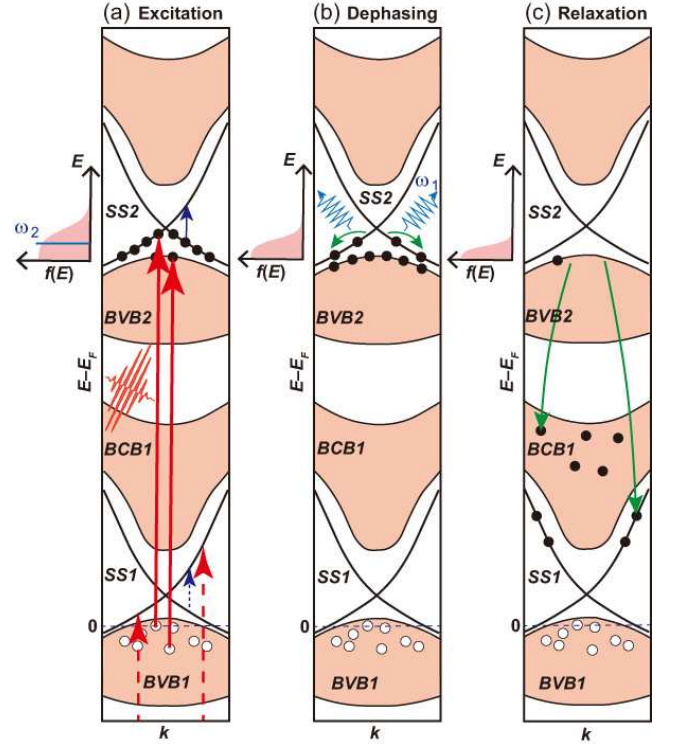}
\caption{(Color online) Schematic band structures for photoexcitation and subsequent relaxation processes in Sb$_2$Te$_3$. (a) A 1.5 eV photon excites electrons from the occupied bulk valence band (BVB$_1$) below the Fermi level to the SS$_2$ band as well as the BVB$_2$ band, generating a non-equilibrium electron population, whose energy distribution $f$($E$) can be approximated using the Fermi-Dirac distribution function. The non-equilibrium electron distribution with a maximum electron temperature of 1800 K ($\approx$ 164 meV) is well overlapped with the optical phonon energy ($\omega_{2}$ as the $A^2_{1g}$ mode), enabling the Fano resonance. The thick dashed arrow represent a possible direct optical transition from deeper-lying bulk states into the SS$_1$ band, whereas the thin solid and dashed arrows represent the electronic transitions within the SS bands at the energy of the $A^2_{1g}$ as a probe action. (b) Ultrafast dephasing of the electrons near the Dirac cone and the coherent emission of the coupled optical phonons ($\omega_{1}$ as the $A^1_{1g}$ mode) within the SS$_2$ bands. (c) Relaxation of electrons from the higher lying BVB$_2$ band into the lower lying bulk conduction band (BCB$_1$) and into the SS$_1$ band is indicated by the arrows. }
\label{Fig5}
\end{figure}
The extracted 1/$q$ shows a strong variance with time delay as shown in Fig. 4(a). To understand the time-dependence of the 1/$q$ value, we have calculated the electron and lattice temperatures using the two-temperature model (TTM). The main idea behind use of the TTM here is that because of the assumption that we are observing electron-phonon thermalization near the Dirac cone, the TTM, which is applicable to the zero-gap metallic systems, can be applied \cite{Kaganov,Allen}. 
The TTM results are presented in Fig. 4(a) and (b)(see Supplemental Material \cite{Sup}). It is found that the TTM predicts the time-dependence of 1/$q$ for time delays $\geq$200 fs, whereas it fails to explain the 1/$q$ dynamics in the early time region ($\leq$200 fs). The 1/$q$ dynamics for $\leq$200 fs can be better fit by a Gaussian function, implying the existence of coherent phenomenon \cite{Prezhdo98,Habenicht06}. The characteristic time of $\approx$200 fs matches the momentum relaxation of the photoexcited electrons in the second Dirac SS (SS$_2$), which was $\leq$165 fs \cite{Soifer17}. The characteristic time of $\approx$200 fs, however, also matches the time scale where the electronic distribution is non-thermal in nature as observed in graphite \cite{Ishida11}. For this non-thermal electron population, the efficient emission of coupled optical phonons is possible at $\leq$200 fs \cite{Ishida11}. We argue that the origin of the Gaussian shape decay is due to the dephasing of the photoexcited electrons near the Dirac cone by the coherent emission of optical phonons.

In addition, the TTM results shown in Fig. 4(b) are useful for understanding the time-lag phenomena observed in Figs. 2 and 3. The electron temperature $T_{e}$ rises immediately after excitation within $\approx$50 fs to $\approx$1800 K, whereas $T_{L}$ rises slowly with a delay time of $\sim$0.5 ps, suggesting a close relationship between the lattice temperature rise and the time-lag. The rise of $T_{L}$ at $\sim$200 fs reflects the nearly coherent emission of optical phonons, and therefore, we propose that the $A^1_{1g}$ mode is generated by coherent emission (see Fig. 5), or equivalently by a thermal gradient force due to the lattice temperature gradient. 
The time-lag becomes less visible for the bulk and 3QL samples due to the fact that the lattice temperature rise becomes much faster for smaller carrier densities (see Supplemental Material \cite{Sup}). Under such conditions, since the carrier density decreases in both the bulk and 3QL samples \cite{Glinka}, a much faster temperature gradient drives the $A^1_{1g}$ mode. Thus, metallic SSs are required to observe the time-lag phenomena. 

The time evolution of the Fano-like line shape can be understood based on Fig. 5. Recently, the existence of an unoccupied SS$_2$ state was discovered \cite{Sobota13,Niesner}. The SS$_2$ is located $\approx$1.5 eV above the lower lying bulk conduction band (BCB$_1$) in $n$-type Bi$_2$Se$_3$ \cite{Sobota13,Niesner}, and plays an important role in the optical interactions of photons in TIs. For Sb$_2$Te$_3$, excited with a 1.5 eV photon energy pulse, optical transitions occur from occupied to unoccupied states \cite{Tu,Wang16,Forster}, as shown in Fig. 5(a). Just after excitation, the photogenerated coherent bulk-phonons ($\omega_{2}$ as the $A^2_{1g}$ mode) and the continuum-like Dirac plasmon of the SS$_2$ strongly couples to each other \cite{Wang16}.
The non-equilibrium electron distribution with a maximum electron temperature of 1800 K ($\approx$164 meV) is well overlapped with the optical phonon energy [see $\omega_{2}$ in Fig. 5(a) for the $A^2_{1g}$ mode], giving rise to quantum interference, resulting in a Fano-like asymmetric line shape. 
Therefore, the transient variation of 1/$q$ is dominated by the loss of coherence of the continuum-like Dirac plasmon excited into the SS$_2$ band. The observed dephasing time ($\approx$200 fs) of the coupling strength (1/$q$) for Sb$_2$Te$_3$ is surprisingly longer than that observed in conventional semiconductors, such as Si ($\approx$50 fs) \cite{Hase03}. 

Unlike in Bi$_2$Se$_3$, where a 1.53 eV photon can promote electrons only to the vicinity of the second bulk valence band (BVB$_2$) \cite{Wang16}, our 1.5 eV photon pump can excite electrons close to the Dirac point of the SS$_2$ of Sb$_2$Te$_3$, because of differing band arrangements \cite{Forster}. Therefore, hot electrons populate the entire SS$_2$ band below the Dirac point, contributing to the formation of a Dirac plasmon. 
Although optical excitation might also promote electrons into higher lying bulk states [Fig. 5(a)], if electron-phonon coupling was present for the SS$_1$ state, a certain minimum time would be required to populate the SS$_1$ state via relaxation from higher lying bulk states [Fig. 5(c)], a phenomena which typically occurs on a $\approx$1 ps time scale \cite{Sobota12}. In the current case on the contrary, we are observing much faster electron-phonon coupling in the form of dynamical Fano-like resonances just after photoexcitation for t$\approx$0, as shown in Figs. 3(b) and 4(a). 
Note that a direct optical transition from deeper-lying bulk states into the SS$_1$ band may be possible in our experiment \cite{Barriga16}, however, the fully populated electrons in the SS$_1$ band would play just a minor role in the observed transient Fano interference, since electronic transitions within the SS$_1$ band are Pauli blocked \cite{Xu17}.
Thus, we are observing a dephasing process [Fig. 5(b)] just after the pumping action. We note that a fluence dependence of the 1/$q$ value has clearly been observed (data not shown), suggesting that we probe mainly the Dirac electron population in the SS$_2$ band. The probing process includes electron continuum transitions available within the SS$_2$, similar to the case of $p$-type Si, where electronic continuum transitions occur between the heavy-hole and light-hole bands \cite{Cerderia73}. 

The observation of the transient Fano interference observed in the present study is comparable to the Fano resonance induced by strong coupling between Weyl fermions and phonons in TaAs \cite{Xu17}, in which a decrease in the 1/$q^{2}$ value was observed upon increasing the temperature. 
We note that similar dynamical Fano-like resonances are also observed in our Bi$_2$Te$_3$ films, indicating it is a general property of topological insulators.

In conclusion, we have investigated the ultrafast coherent optical phonon dynamics in ultrathin films of Sb$_2$Te$_3$ using a time resolved optical pump-probe technique. 
The discrete wavelet transformed phonon spectra for different time scales exhibit a Fano-like asymmetric line shape, which is attributed to quantum interference between a continuum like coherent Dirac plasmon and the $A^2_{1g}$ phonon. 
The inverse of the Fano asymmetry parameter decays within 1 ps, and its time dependence can be well fit to a electron temperature calculated by the two-temperature model, whereas the TTM cannot well explain the Gaussian decay profile observed up to $\approx$200 femtoseconds after photoexcitation. The Gaussian decay observed in the early time region may be a result of a non-thermal electron population, which cannot be described by a Fermi-Dirac distribution. In addition, we have found the observed time lag between the two optical modes can be reasonably ascribed to a different thermal gradient force due to the delayed lattice temperature gradient. Our findings may open up a novel route for investigating quasiparticle dynamics in topological insulator materials and may be also important for applications in optical switching and sensing devices based on topological insulator materials. 

\begin{acknowledgments}
This research was financially supported by CREST (NO. JPMJCR14F1), JST, Japan and JSPS KAKENHI-17H02908, MEXT, Japan. 
We acknowledge Ms. R. Kondou for sample preparation. 
\end{acknowledgments}

\pagebreak


\begin{thebibliography}{99}
\bibitem {note1}
We have in fact grown 3, 4, 5, 6, 7, 8, 10, and 50 QL samples and have measured all the samples to investigate the thickness dependence. 
The reason why we show only the 8 QL sample in the main text is that the 8 QL exhibited a more prominent Fano-like line shape, and therefore 
we focused on it in the present paper.
\bibitem {note2}
The average size of single domain of the highly oriented polycrystalline sample is $\approx$100 nm, which is much smaller than the beam spot diameter of $\approx$70 $\mu$m. The optical penetration depth ($\approx$14 nm) is, however, nearly an order of magnitude smaller than the grain size, supporting negligibly small contribution from trivial edge terminations along the domain wall.
\end{thebibliography}
\end{document}